\documentstyle[preprint,aps]{revtex}
\begin{document}
\title{Influence of parallel magnetic fields on a single-layer two-dimensional
electron system with a hopping mechanism of conductivity}
\author{I. Shlimak$^{1,2}$, S. I. Khondaker$^{1}$, M. Pepper$^{1}$, and D. A. Ritchie%
$^{1}$}
\address{$^{1}$Cavendish Laboratory, University of Cambridge, Madingley Road,\\
Cambridge CB3 0HE, UK\\
$^{2}$Jack and Pearl Resnick Institute of Advanced Technology,\\
Department of Physics, Bar-Ilan University, Ramat-Gan 52900, Israel}
\maketitle

\begin{abstract}
Large positive (P) magnetoresistance (MR) has been observed in parallel
magnetic fields in a single 2D layer in a delta-doped GaAs/AlGaAs
heterostructure with a variable-range-hopping (VRH) mechanism of
conductivity. Effect of large PMR is accompanied in strong magnetic fields
by a substantial change in the character of the temperature dependence of
the conductivity. This implies that spins play an important role in 2D VRH
conductivity because the processes of orbital origin are not relevant to the
observed effect. A possible explanation involves hopping via double occupied
states in the upper Hubbard band, where the intra-state correlation of spins
is important.
\end{abstract}

\bigskip 

Investigation of the two-dimensional (2D) conductivity of localized
electrons in a GaAs/AlGaAs heterostructure \cite{Tremblay90,Keuls,Saiful-1}
showed that at low temperatures and in zero magnetic field, the temperature
dependence of the longitudinal resistivity $\rho _{xx}(T)$ has the
variable-range-hopping (VRH) form: 
\begin{equation}
\rho (T)=\rho _{0}\exp \left[ \left( T_{0}/T\right) ^{x}\right] ,
\label{eq-1}
\end{equation}
where $\rho _{0}$ is a prefactor, $T_{0}$ is a characteristic temperature,
and the exponent $x$ depends on the shape of the density-of-states (DOS)
near the Fermi level. The exponent $x=1/2$ corresponds to the existence of a
soft (linear) Coulomb gap in the DOS near the Fermi level \cite
{Pollak70,Knotek,ES75,Book}. This is called Efros-Shklovskii (ES) VRH
conductivity. The exponent $x=1/3$ (Mott VRH conductivity) corresponds to a
constant DOS at the Fermi level \cite{Mott68}. Mott VRH can be observed if
the influence of the Coulomb interaction is negligible, for example, at high
temperatures when the optimal hopping energy exceeds the width of the
Coulomb gap \cite{Saiful-2}, or at low temperatures in gated samples, when
the hopping distance became larger than twice the distance to the metallic
gate, which results in screening of the Coulomb interaction \cite{Aleiner}.

In 3D conductivity, hopping magnetoresistance (MR) has been studied in a
large number of publications. The theory of positive hopping
magnetoresistance (PMR) developed by Shklovskii \cite{Book} is based on the
orbital shrinkage of the electron wavefunction in a magnetic field, causing
a reduction in the overlap between states and quadratic field dependence of
the logarithm of resistivity:

\begin{equation}
\rho \left( B,T\right) =\rho \left( 0,T\right) \exp [B^{2}/B_{0}^{2}(T)]
\label{eq-2}
\end{equation}
Here, $B_{0}$ is a parameter that depends on temperature, $B_{0}\propto
T^{m} $, $m=3x$, i.e. $m=3/2$ for ES\ VRH conductivity.

In 2D, measurements of MR in the VRH regime have been reported for both
magnetic field \ orientations: perpendicular and parallel to the 2D plane.
In weak fields, the effect of negative magnetoresistance (NMR) was observed
and explained by Nguyen {\it et al}. \cite{Nguyen} and by Sivan {\it et al}. 
\cite{Sivan} as due to quantum interference of different hopping paths
between initial and final states. Ye {\it et al}. \cite{Ye} reported PMR in
the high-field limit for both orientations and NMR in the low-field limit
for the perpendicular direction. Their sample contained 20 parallel Si-$%
\delta $-layers embedded in GaAs. The PMR in both orientations was explained
as being due to orbital shrinkage of the electron wavefunction. Raikh {\it %
et al.} \cite{Raikh-1} observed a similar effect in $\delta $-doped GaAs
with 18 parallel layers. The nonmonotonic NMR in the perpendicular
orientation\ has been explained as being due to a combination of an
interference mechanism, which suppresses the effect of weak localization in
a magnetic field, and an incoherent mechanism, caused by the influence of
the magnetic field on the DOS at the Fermi level which defines the magnitude
of the Mott VRH parameter $T_{0}$ \cite{Raikh-2}. Similar results were also
reported for a parallel magnetic field by D\"{o}tzer {\it et al}. \cite
{Klaus} for a double layer system.

It should be mentioned that so far, hopping MR in parallel magnetic fields
has been observed in multilayered systems. Moreover, in Ref. \cite{Chehi} it
was emphasized that 2D MR is not quite negligible only in the case where the
structure contains two or more parallel 2D subsystems that are a tunneling
distance apart. In this paper, we report the first measurements of MR in the
VRH regime in parallel magnetic fields in a single-layer 2D electron system.
We observed a large PMR accompanied in strong fields by a substantial change
in the character of the temperature dependence of the conductivity.

The samples investigated were fabricated from the wafer used in our previous
study, Ref. \cite{Tremblay90,Saiful-1}, where full details of the layer
compositions and doping are given. An essential feature of the sample
structure is the presence of a delta-doped layer on the AlGaAs side of the
electron gas, 0.6 nm away from the heterojunction. The samples were
patterned into $80\times 720~\mu $m Hall bars, and the as-grown 2D carrier
density and low-temperature mobility were $n=4.65\times 10^{11}$~cm$^{-2}$
and $\mu =4.5\times 10^{4}$~cm$^{2}$/Vs. The carrier density $n$ was varied
by the application of a negative gate voltage $V_{g}$ to a surface gate that
is $d=90~$nm above the heterojunction; at low temperatures, the electron gas
is fully depleted at $V_{g}=-0.70$~V. MR measurements were carried out in
both a $^{3}$He cryostat and in a dilution refrigerator. In the $^{3}$He
cryostat, the sample could be aligned to better than 0.1$^{0}$ with the
parallel magnetic field by using an {\it in-situ} rotation mechanism. In the
dilution refrigerator, the sample was glued to the probe and therefore was
susceptible to misalignment. After temperature stabilization, the
resistivity (the resistance per square) was measured from the Ohmic part of
the four-probe DC $I$-$V$ characteristics. At higher carrier densities, a
four-terminal AC technique was also used with 1 nA current at a frequency of
2-4 Hz. The measurements in fields parallel to the 2D plane correspond to
the current being parallel to the field. We have also measured the case
where the magnetic field is parallel to the 2D plane but perpendicular to
the current. No anisotropy was observed.

Figure 1(a) shows the resistivity $\rho (B)$ normalized by the zero-field
resistivity $\rho (0)$ versus the magnetic field for several carrier
concentrations at $T=0.3$ K. It is seen that $\rho (B)/\rho (0)$ increases
with magnetic field. The increase is stronger at lower electron densities.
At the carrier density $n=9.18\times 10^{10}$ cm$^{-2}$, the increase is
almost seven-fold at a field of 10 Tesla. Figure 1(b) shows $\rho (B)/\rho
(0)$ at different temperatures for a fixed carrier density $n=9.18\times
10^{10}$ cm$^{-2}$. It is seen that the increase is stronger at lower
temperatures.

Thus, the effect of PMR is stronger for lower densities and lower
temperatures, which corresponds to ES VRH conductivity ($x=1/2)$ \cite
{Saiful-1}. At higher densities and higher temperatures, which corresponds
to Mott VRH conductivity $(x=1/3),$ the effect of PMR is much weaker.
Moreover, a negative effect (NMR) can be seen in the low-field regime ($B<$
2 T). Measurements of Hall voltage at fixed $n$ and low $B$ show that there
is a misalignment of $\sim 1^{0}$ with the parallel magnetic field. To check
the existence of NMR in parallel fields, we measured samples in a $^{3}$He
cryostat having an {\it in-situ} rotation mechanism. By monitoring the Hall
voltage, we could align the sample to better than 0.1$^{0}$ with the
parallel field. These measurements showed that there is no NMR in parallel
fields at all temperatures. The $^{3}$He cryostat\ was supplied by a
superconducting magnet with relatively weak magnetic field, therefore all
data in strong fields were obtained in the dilution refrigerator.\ Weakness
of the PMR and entanglement with NMR in the case of Mott VRH leads one to
stress ES VRH in further discussion, while the influence of a parallel
magnetic field on the Mott VRH is qualitatively the same.

To determine the functional form of $\rho (B),$ we plot in Fig. 2 the values
of $\ln [\rho (B)/\rho (0)]$ versus $B$ on a log-log scale for carrier
density $n=9.52\times 10^{10}$ cm$^{-2}$. It can be seen that at higher
temperatures, the data follow a $B^{2}$-law, Eq. (\ref{eq-2}) up to $B=8$ T
(deviations at low fields reflect the interference with NMR), whereas at low
temperatures, the $B^{2}$-law is observed only at fields up to 2-3 T. For $%
B>4$ T, a weaker field dependence is valid. Similar behavior has been
observed for other carrier densities. From $B^{2}$-dependencies, one can
determine the parameter $B_{0}$. The inset to Fig. 2 shows $B_{0}$ vs. $T$
on a log-log scale. The straight line corresponds to $B_{0}^{2}\propto T^{m}$%
, where $m=1.2$ is a little bit less than the predicted value $m=1.5$ for 3D
ES\ VRH.

Let us now discuss the temperature dependence of the resistivity $\rho (T)$
in fixed magnetic fields and carrier density. Figure 3(a) shows the
resistivity at different magnetic fields versus $T^{-1/2}$ for $n=9.52\times
10^{10}$ cm$^{-2}.$ At zero field, the low-temperature part of the curve
fits the straight line, which shows that the conductivity obeys the ES\ VRH
law, Eq. (\ref{eq-1}) with $x=1/2.$ It is remarkable that in our delta-doped
GaAs/AlGaAs heterostructure, the hopping prefactor $\rho _{0}$ in ES\ VRH is
temperature-independent at zero field, and equal to $(h/e^{2})=25.8$ k$%
\Omega $ (Fig. 3(a), see also Ref. \cite{Saiful-1}). The same
temperature-independent prefactor for ES VRH at zero field was previously
observed in a different material, Si-MOSFET \cite{Mason} in a narrow
interval of electron densities near the metal-insulator transition. This was
understood in Refs. \cite{Saiful-1,Mason} as a hint that the VRH
conductivity at zero magnetic field at given electron densities is not
determined by the conventional phonon-assisted scattering, because in this
case, the prefactor must be material- and temperature-dependent \cite{Book}.
As a possible alternative, a mechanism of hopping conductivity via
electron-electron scattering, first discussed by Fleishman {\it et al}. \cite
{Fleishman}, was suggested in Ref. \cite{Aleiner}. Comparison of the
low-temperature conductivity data in different GaAs/AlGaAs heterostructures 
\cite{Keuls,Saiful-1} leads to the conclusion that existence of a
delta-doped layer in the close vicinity of the 2D conducting plane favors
the conductivity mechanism with universal prefactor $\rho _{0}=(h/e^{2})$.
Indeed, the VRH in GaAs/AlGaAs heterostructures without additional
delta-doped layer is characterized by a temperature-dependent prefactor of
the form $\rho _{0}=AT^{m},$ $m=0.8$ or 1.0 \cite{Keuls,Ebert,Briggs} which
is peculiar for the conventional phonon-assisted mechanism of VRH
conductivity.

For analysis of $\rho (T)$-dependencies, it is usful to plot the derivative
of the curves $w=-\partial (\ln \rho (T))/\partial (\ln T)=x(T_{0}/T)^{x}$
versus $T$ on a log-log scale, because the slope of the lines on this scale
gives directly the value of index $x$ in Eq. (\ref{eq-1}). The result of
this analysis is presented in Fig. 4. One can see that at zero field and low
temperatures (below 1 K), the power $x$ in Eq. (\ref{eq-1}) is indeed equal
to 1/2, which corresponds to ES VRH. Increasing $T$ leads to a crossover to
Mott VRH ($x=$ 1/3), caused by the increase of the optimal hopping band over
the width of the Coulomb gap \cite{Saiful-2}. Figure 4 shows that in strong
magnetic fields 6 and 8 T, the value of $x$ at low temperatures increases up
to $x\approx 0.8$. The same value of $x$ was obtained for $B=10$ T. There is
no theoretical justification for VRH conductivity with $x=0.8.$ Therefore,
one suspects that a stronger temperature dependence of the resistivity is
caused by the temperature-dependent prefactor. Following \cite{Keuls},\ we
assume that the prefactor becomes temperature-dependent of the form $\rho
_{0}=AT^{4/5},$ and plot ln($\rho _{xx}/T^{4/5})$ versus $T^{-1/2} $ in Fig.
3(b). It is seen that on this scale, the low-temperature data for $B=$ 6, 8
and 10 T again fit to a $T^{-1/2}$-behavior. This result could be
interpreted as the restoration of the conventional phonon-assisted hopping
mechanism in strong parallel magnetic fields. It would be interesting to
continue these measurements at higher magnetic fields to answer the question
of whether or not $x$ saturates at the value of $0.8.$

In this experiment, the processes of orbital origin are not relevant to the
observed effect, because the magnetic field, applied parallel to the single
2D conducting layer of electrons, couples only to the electron spins. This
implies that spins play an important role in 2D hopping conductivity.
Meanwhile, there is no reason\ for this effect unless the doubly-occupied
states of the upper Hubbard band are not involved in the hopping motion. The
two electrons at the site should have opposite spins. Thus, the spin
alignment by the magnetic field decreases the occupation numbers for the
upper Hubbard band, which leads to PMR. For 3D VRH conductivity, the
corresponding mechanism was first considered by Kurobe and Kamimura \cite
{Kamimura}. Agrinskaya and Kozub \cite{Kozub}, analysing the data on
magnetic-field dependence of activation energy for 3D nearest-neighbour
hopping conductivity have concluded that in many occasions it is related to
the upper Hubbard band. Then, they also observed the features predicted by
Kamimura {\it et al}. in VRH conductivity which evidences an admixture of
the upper Hubbard band states at the Fermi level.

One should realize that the localization length in the upper Hubbard band is
much larger than for the lower band making the upper band dominant for the
hopping process. Moreover, it can in principle allow delocalization, even if
the states in the lower Hubbard band are strongly localized. Note that with
an account of energy dependence of the density-of-states in the upper
Hubbard band and greater degree of delocalization, the temperature
dependence of VRH appears to be more complex as predicted by Kamimura {\it %
et al}. and can lead to the activation-like law \cite{Agrinskaya}.

This work was supported by the UK Engineering and Physical Science Research
Council. The authors thank A. L. Efros, B. I. Shklovskii, and V. I. Kozub
for fruitful discussions. SIK acknowledges support from the Commonwealth
Scholarship Commission in the United Kingdom.


\begin{references}
\bibitem{Agrinskaya}  N. V. Agrinskaya and V. I. Kozub, private
communication.\newpage

{\large Figure captions\bigskip }

Fig. 1. Normalized resistance $\rho (B)/\rho (0)$ as a function of magnetic
field $B$ for different electron densities $n$ at fixed temperature $T=300$
mK (a) and for different temperatures at fixed $n=9.18\times 10^{10}$ cm$%
^{-2}$ (b).\bigskip

Fig. 2. Magnetoresistance $\ln [\rho (B)/\rho (0)]$ \ for $n=9.52\times
10^{10}$ cm$^{-2}$ versus $B$\ on a log-log scale. The solid and the dashed
lines correspond to $\ln \rho \propto B^{2}$ and $\ln \rho \propto B^{1.4}$,
respectively. The insert shows $B_{0}^{2}$ vs. $T$ on a log-log scale. The
solid line corresponds to $B_{0}^{2}\propto T^{1.2}.\bigskip $

Fig. 3. Temperature dependence of $\rho _{xx}$ (a) and $\rho _{xx}/T^{0.8}$
(b) versus $T^{-1/2}$ for $n=9.52\times 10^{10}$ cm$^{-2}$ at $B=0,6,8$ T.
The straight lines are guides to the eye.\bigskip

Fig. 4. The derivative $w=-\partial (\ln \rho (T))/\partial (\ln
T)=x(T_{0}/T)^{x}$ as a function of $T$ on a log-log scale for $B=0,6,8$ T, $%
n=9.52\times 10^{10}$ cm$^{-2}$. The exponent $x$\ is given by the slope of
the curve. The solid lines show different temperature dependences of the
resistivity $\ln \rho \propto T^{-x}.$
\end{references}
\end{document}